\documentclass[ ]{aastex631}

\usepackage{graphicx} 
\usepackage{enumitem}
\usepackage{color}
\usepackage{longtable}
\usepackage{float}
\graphicspath{{./}{figures/}}
\usepackage{topcapt}
\shorttitle{Gravity modes in RR\,Lyrae}
\shortauthors{Merieme Chadid}
\begin{document}

\date{Received 2021 June 24; Accepted 2021 November 4}
\title{Detection of Gravity Modes in RR Lyrae Stars \footnote{Contact: chadid@unice.fr}} 

\author{Merieme Chadid}
\affiliation{University of C\^ote d'Azur, Nice Sophia--Antipolis University, CNRS--UMR 7250, CS 34229, 06304 NICE Cedex 4, France }

\begin{abstract}

We report the detection of  gravity modes in RR\,Lyrae stars. Thanks to  PAIX, the first Antarctica polar photometer. Unprecedented and uninterrupted  $UBVRI$  time--series photometric ground--based data are collected during  150 days from the highest plateau of Antarctica. PAIX light curve analyses reveal an even richer power spectrum with mixed modes in RR\,Lyrae stars. A nonlinear nature of several dominant peaks, showing lower and higher frequencies, occur around the dominant fundamental radial pressure mode. These lower frequencies and harmonics linearly interact with the dominant fundamental radial pressure mode and its second and third overtone pressure modes as well. Half--integer frequencies are also detected, likewise side peak structures  demonstrating that HH\,puppis is a bona--fide Blazhko star. Fourier correlations are used to derive underlying physical characteristics for HH\,puppis. The most striking finding is the direct detection of  gravity waves. We interpret the excitation mechanism of  gravity waves in RR Lyrae stars by the penetrative convection driving mechanism. We demonstrate that RR\,Lyrae stars pulsation is excited by several distinct mechanisms, and hence RR Lyrae stars are simultaneously $g$ modes and $p$ modes pulsators. Our discoveries make RR\,Lyrae stars very challenging stellar objects, and provide their potential to undergo at the same time $g$ modes and $p$ modes towards an advancement of theory of stellar evolution and a better understanding of the Universe.

\end{abstract}

\keywords{observational astronomy: Ground-based astronomy -- astronomical instrumentation: photometer --  observation methods -- astronomy data analysis -- hydrodynamics -- stellar convection envelopes -- stellar pulsations -- pulsating variable stars: RR Lyrae. }

\section{Introduction}\label{Intro}
Stellar Pulsation and Asteroseismology, the real music of the star, are a good clue towards an understanding of internal structure and dynamics of the stars from oscillations observed at their surface. Different modes penetrate to different depths in the star. Mathematically, there are two main sets of solutions to the equation of motion for an oscillating star, leading to two types of pulsation modes: the $p$ modes, acoustic waves, pressure is the primary restoring force for a star perturbed from equilibrium. The $p$ modes are most sensitive to conditions in the outer part of the star. The $g$ modes, gravity waves, buoyancy is the restoring. They are most sensitive to conditions in the deep interior of the star. In principle, the $g$ modes  have finite amplitude in the outer part of star and are observed on the surface as well. Otherwise, in massive main-sequence stars, the $g$ mode are confined outside the convective core. The major argument for years consisted in the big challenge in finding the $g$ modes and providing a possible mechanism of their excitation. \\

RR\,Lyrae stars are considered to be the classical radial pulsators, oscillating in $p$ modes. Most of them are monoperiodic stars with an oscillation period near half a day. They are of enormous cosmological and galactic importance. Extreme Population\,II, RR\,Lyrae stars were discovered first in globular clusters by Bailey in 1895. They are low mass stars and are used to estimate the distance and the age of the clusters. Known as standard candles of galactic evolution, RR\,Lyrae stars have been observed for more than a century and subdivided into  two Bailey classes RR\,ab and RR\,c stars (\cite{Bailey02}). Such classification is based on the skewness and amplitude of the light curve, and the pulsation period as well. A fourth class, RR\,d was introduced later, oscillating in both the fundamental and first overtone. Spectroscopy studies demonstrated that RR\,Lyrae stars are affected by nonradial oscillations as well (\cite{Chadid99}). Such stars are also affected by hypersonic shock waves (\cite{Chadid08}). \\

However, the big challenge is the Blazhko effect (\cite{Blazhko07}). More than a century after its discovery, the Blazhko effect still remains a mystery. That greatly complicates the accurate measurements of the atmospheric dynamics of RR Lyrae stars and introduces unavoidable complications in an understanding of stellar pulsation and evolution. The most popular model, trying to explain the Blazhko effect, predicts the
dependence of the Blazhko amplitude upon the strength of a magnetic field of the order
of 1.5 kG --the oblique--dipole rotator model-- (\cite{Shibahashi95}). Having covered several pulsation periods and Blazhko cycles spread over a period of 4--years with high precision longitudinal magnetic field measurements, \cite{Chadid04} reported no detection of magnetic fields in RR\,Lyrae stars, and concluding that no magnetic model is able to explain the Blazhko effect, and further understanding of the origin of the Blazhko effect still needs new observations. Recently, \cite{Stothers10} has proposed a new explanation of the Blazhko effect caused by changes in the structure of the outer convective zone, generated by an irregular variation of the magnetic field. A resonance between a radial mode and a nonradial mode hypothesis (\cite{Now01} and  \cite{Dzi04}), and a radial mode resonance hypothesis (\cite{Buch11}) were also suggested to explain the Blazhko effect. \\

Up to now, the best applications in  RR\,Lyrae stars study have relied dominantly on classical ground--based data sets (\cite{Alcock03} and  \cite{sos19}), improved progressively by use of space telescopes such as CoRoT (\cite{Auvergne09} and \cite{Chadid12}), Kepler (\cite{Gilli10}, \cite{benko14} and \cite{molnar15} ) and TESS (\cite{Molnar21}), and accomplished by  implementing a new way, Antarctica polar observations  with PAIX long uninterrupted and continuous precision observations over 150 days from the ground, 1 polar night, and without the regular interruptions imposed by the earth rotation (\cite{Chadid19} and \cite{Chadid14}).\\

The paper is organised as follows. In Section \ref{data} we briefly describe the observations and the data reduction process. The description and application of methods used for the frequency analysis are presented in Section \ref{frequency}. The fundamental parameters, shock waves and dynamics are discussed in Section \ref{parameters} and Section \ref{shock}. In Section \ref{gmodes}, we demonstrate the $g$ mode detection and we provide an explanation. Finally, some concluding remarks and our future plans are given
in Section \ref{con}.

\section{Observations and data reduction}\label{data}

\subsection{PAIX --First Robotic Antarctica Photometer--}
PAIX --Photometer AntarctIca eXtinction--  the first robotic multi--band photometer (\cite{Chadid18}),  has been antarctized and robotized to run under extreme weather and human conditions with temperatures as low as --83$^{\circ}$C, altitude higher than 4000\,m according to the low pressure at latitude\,=\,-75\,$\deg$ in the heart of Antarctica. PAIX is attached to the Cassegrain focus of a 40--cm Ritchey--Chretien optical telescope, with a F/D ratio of 10, supported
by an equatorial mount AstroPhysics 1200 and operates in the open field, without any shelter and installed at ice level. At the focus are successively installed a robot--focus (Optec TCF), a filter wheel (SBIG--CFW6) equipped with Johnson-Cousin filters and a CCD dual chip (SBIG--ST10--XME), all included in a thermalized box. Collecting simultaneously multicolor light curves of several targets within the same 12.2 x 8.2 arcmin field of view. All PAIX components are remotely controlled and setup through PAIX ACquisition Software --PACS-- implemented on
PAIX experiment in Antarctica, accessible from anywhere through a virtual private network (VPN) and a remote desktop. For more technical details on the PAIX multi--band photometer and data we refer to the invited SPIE paper \cite{Chadid16}.

\begin{figure*}[ht!]
\plotone{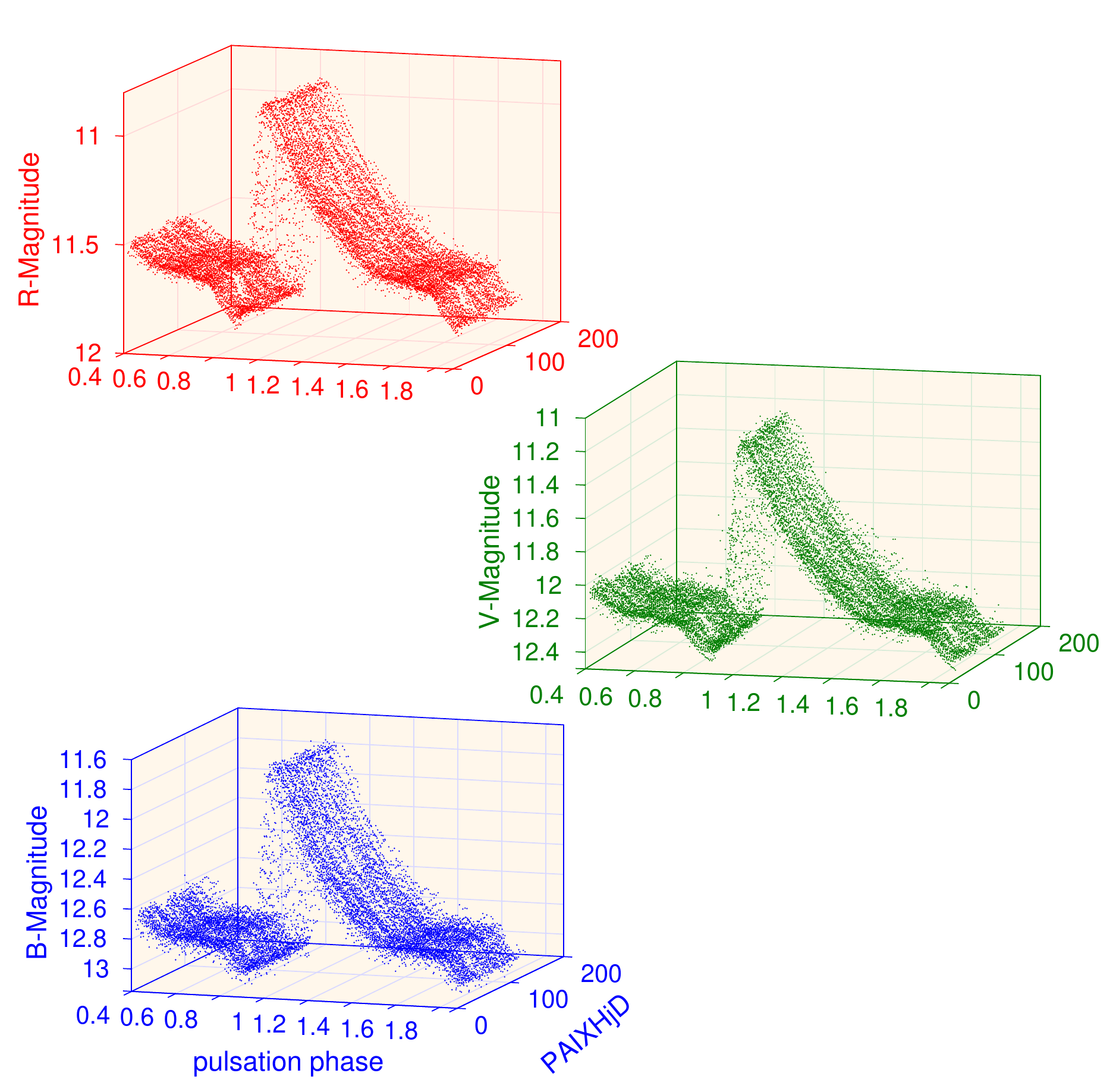}
\caption{Three-dimensional observed PAIX light curve of RR Lyrae star HH Puppis folded with the pulsation period (0.390\,d), during the observing run of 2011 polar wintertime. Axis refer to pulsation phase, magnitude and PAIX Heliocentric Julian Day. \label{fig}}
\end{figure*}

\subsection{PAIX Observations}
The PAIX data used in this study are the calibrated light curves from the magnitude extraction that we measure by PAIX CCD during the observing run starting on 12 May 2011 and ending on 25 September 2011.  The 17 500 CCD images of the target, HH\,Puppis, 11.31\,V-mag, with J2000 $\alpha$ right ascension  $07^{\mathrm h}$ $20^{\mathrm m}$ $35^{\mathrm s}{55}$ and $\delta$ declination $-46^{\mathrm h}$ $42^{\mathrm m}$ $30^{\mathrm s}{11}$, have been collected throughout the UBVRI color filters  with an exposure time $\Delta$t $\leq$ 60\,s, and processed even when the sun is at 8$^{\circ}$ below horizon. In this case, the accumulated observable time spans, without any interruption, over  136 days of 2011 polar wintertime.  We use for our purpose the V--filter passband in the Bessel Johnson--Cousins system (\cite{bes05}) stands for an effective wavelength midpoint $\lambda_{eff}$\,=\,540\,nm and Full width at half maximum $\Delta\lambda$\,=\,100\,nm. Such spectral bandpass is 4.5 times narrower than the Kepler bandpass which is from 400\,nm to 850\,nm (\cite{Koch10}).

\subsection{PAIX data processing pipeline --PPP--}

The PAIX Pipeline Package --PPP-- provides  work flows for processing PAIX observed polar photometric data.  Such software tool has been developed and maintained  to  provide measurements of both  magnitude and extinction extractions and their precision towards Light Curves and Bouger's lines simultaneously (\cite{Chadid19}). The  PAIX Pipeline --PPP-- is principally based on a set of four softwares, (1)  raw UBVRI images are corrected from dark, bias, flat fields and then corrected for hot pixels. The measurements of the magnitude of each star is performed with use of Sextractor, (2) the coordinates of the target and the position of a given known reference star are determined with the target  magnitude 
precision, (3)   the previous processing is used to plot light curves and finally (4) the magnitude variation of the reference star as a function of the air-mass is analyzed to compute extinction coefficient, according to Bouguer's theory, and plots  Bouguer's lines.

\section{Detection of frequencies and results}\label{frequency}

Figure\,\ref{fig} shows the PAIX HH\,Puppis light curve  folded with the pulsation period $0.390\,d$ over 136 days. The PAIX data of HH\,Puppis provide 349 consecutive pulsation cycles and a homogeneous and uninterrupted coverage.\\

A frequency analysis of the photometric data was performed using Period04 (\cite{Lenz05}) and PDM (\cite{Zal14}) . Both softwares perform a projection of the discrete photometric signal on a trigonometric basis and all led to the same results with a slight difference at higher orders. In this study, we present the results of the Period04 frequency analysis. The Fourier decomposition consists of fitting
the magnitude measurements by means of the series
\begin{equation}
m(t) = A_0 + \sum_{i=1, N}A_i\sin[2 \pi (F_i(t - T_0) + \Phi_i)],
\end{equation}
where $T_{0}$ is the initial epoch value of the data set PAIXJD  $T_{0}$ = 245569 and  $\Phi$ the normalized phase.
\subsection{Fundamental radial pressure mode and harmonics}\label{}
The original spectrum is dominated by the main pulsation frequency  $f_{0}$ = 2.559196\,$d^{-1}$ with 
a $\sigma$$_{f_{0}}$ = $1.485.10^{-6}$\,$d^{-1}$, and its harmonics up to the 30th order (fig.\,\ref{frequency}). Figure\,\ref{order1} demonstrates the strong nonlinearity of the fundamental radial mode showing that the harmonic main frequency amplitude ratio follows a strong  exponential variation along the harmonic order. However, the  high orders fails to comply with this scenario. a standstill appears at the $12^{th}$ order. The same phenomenon was detected and interpreted in  the CoRoT RR\,Lyrae stars (\cite{ChadidCorot} and \cite{Benko16}). 	

\subsection{Gravity modes and harmonics}\label{}
\begin{figure*}
\gridline{\fig{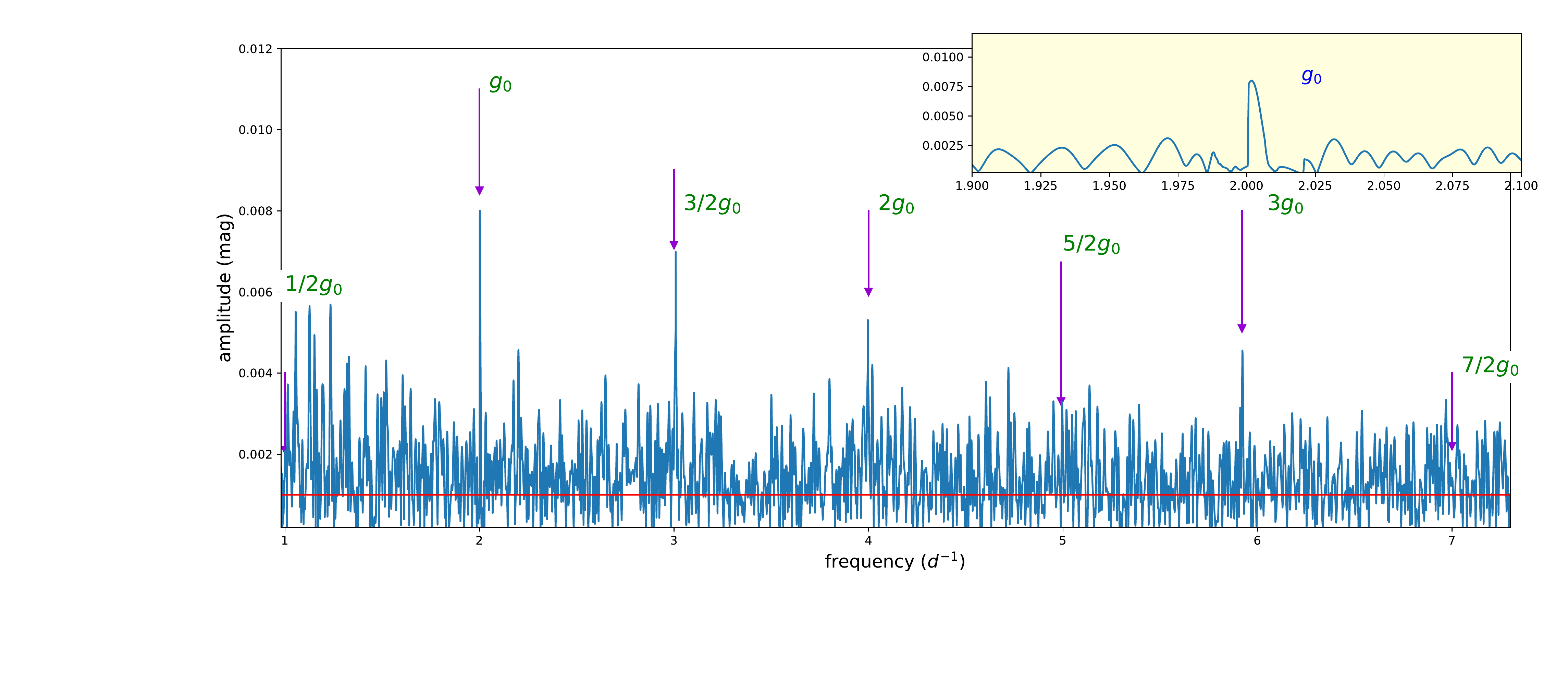}{1.0\textwidth}{(a)}}
\gridline{\fig{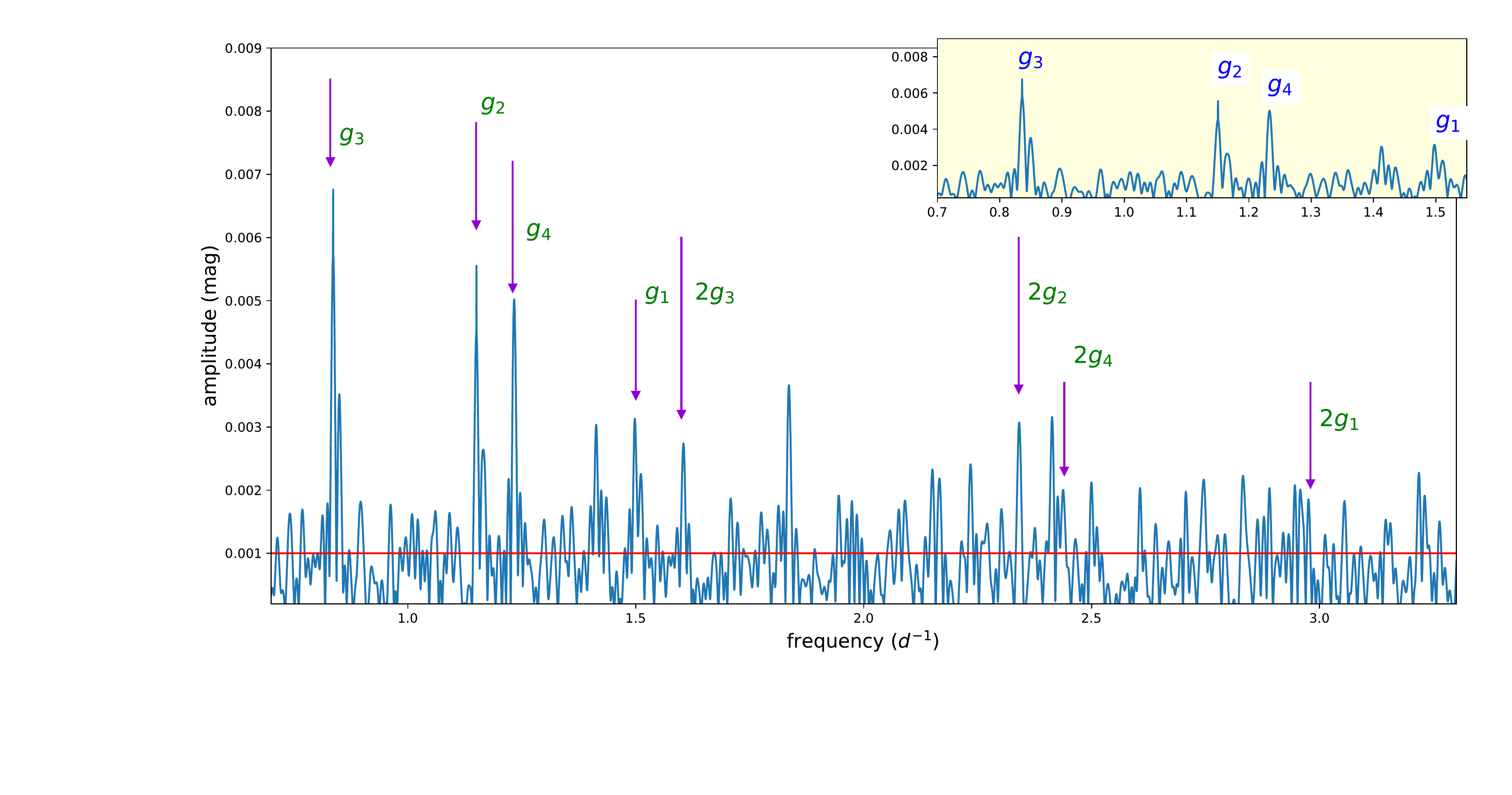}{1.0\textwidth}{(b)}}
\caption{From the upper to lower panels: (a) the amplitude spectrum of the PAIX light curve of RR Lyrae star HH Puppis after prewhitening the fundamental radial frequency $f_{0}$ and its harmonics.  The insert panel shows the nonradial gravity frequency $g_{0}$, and (b) after prewhitening the fundamental radial frequency, its harmonics, the nonradial gravity frequency $g_{0}$ and its harmonics. The insert panel shows the nonradial gravity frequencies  $g_{1}$, $g_{2}$,  $g_{3}$ and $g_{4}$. \label{gmode}} 
\end{figure*}

After prewhitening the spectrum with the main pulsation frequency, and its harmonics, a dominant peak is detected at the frequency $g_{0}$ = 2.001301\,$d^{-1}$ with a $\sigma$$_{g_{0}}$ = $1.523.10^{-5}$\,$d^{-1}$, and its harmonics at 5th order. Significant frequencies $g_{3}$ = 0.835433\,$d^{-1}$$\pm$0.00002, $g_{2}$ = 1.150662\,$d^{-1}$$\pm$0.00002, $g_{4}$ = 1.233302\,$d^{-1}$$\pm$0.00002 and $g_{1}$ = 1.498787\,$d^{-1}$$\pm$ 0.00002 are detected with harmonics at second order (fig.\,\ref{gmode}). A detailed analysis reveals a complex structure in the frequency domain around the dominant peak $g_{0}$,  $g_{3}$, $g_{2}$, $g_{4}$ and $g_{1}$. All these frequencies are interacting with the fundamental frequency $f_{0}$. 
Such  combination terms between $g_{0}$, $g_{3}$, $g_{2}$, $g_{4}$, $g_{1}$ and the fundamental radial frequency $f_{0}$ exclude the possibility that $g_{0}$, $g_{3}$, $g_{2}$, $g_{4}$, $g_{1}$ are actually related to a back--ground contact system, measured with HH\,Puppis in the PAIX window. The list of detected frequencies is given in Table\,\ref{frequency}. It is a great challenge to find an answer to the origin of $g_{0}$, $g_{3}$, $g_{2}$, $g_{4}$ and $g_{1}$. The rotation period, tidal effects and higher-order radial overtones pressure modes can be excluded as an explanation for the occurrence of those frequencies. Those frequencies seem  to be too short to be the rotation period of the RR\,ab star HH\,Puppis (\cite{Preston19}) or  be connected to the orbital period. We could hardly explain the combination terms in this framework. Those frequencies are lower than the fundamental radial frequency ${f_{0}}$, higher--order radial overtones radial pressure modes can be easily excluded as well.\\

The most striking features are that these frequencies are lower and show  frequency ratios to the fundamental frequency $f_{0}$ of  $g_{0}/f_{0}$ = 0.78,  with a period ratio $f_{0}$/$g_{0}$ to the main pulsation period $P_{0}$ of 1.28, $g_{1}/f_{0}$ = 0.58,  with a period ratio $f_{0}$/$g_{1}$ to the main pulsation period $P_{0}$ of 1.72, $g_{2}/f_{0}$ = 0.45,  with  a period ratio $f_{0}$/$g_{2}$ to the main pulsation period $P_{0}$ of 2.22, $g_{3}/f_{0}$ = 0.33,  with  a period ratio $f_{0}$/$g_{3}$ to the main pulsation period $P_{0}$ of 3.03, and $g_{4}/f_{0}$ = 2.22,  with a period ratio $f_{0}$/$g_{4}$ to the main pulsation period $P_{0}$ of 2.08. 
We explain $g_{0}$, $g_{3}$, $g_{2}$, $g_{1}$ and ${g_{4}}$ as gravity modes. Such gravity modes must be nonradial modes since for radial oscillations, the perturbation in the gravitational field is eliminated analytically (\cite{Chris98}).

Table\,\ref{tabratio} gives the amplitude ratio between the fundamental main frequency and the gravity modes. This table  shows that  the nonradial gravity mode $g_{0}$ is the dominant pulsation mode after the main fundamental frequency  $f_{0}$ and is characterized by a nonlinear behaviour as shown by the harmonic gravity mode amplitude ration evolution (fig\,\ref{order1}) and the pronounced nonsinusoidal shape of its light curve in fig\,\ref{LC}. The additional frequencies 
1/2 $g_{0}$ = 1.0006743\,$d^{-1}$, 3/2 $g_{0}$ = 3.001673\,$d^{-1}$, 5/2 $g_{0}$ = 5.003268\,$d^{-1}$ and  7/2 $g_{0}$ = 7.004667\,$d^{-1}$ might be the  half--integer frequencies of the nonradial  gravity  mode $g_{0}$. The third half--integer frequency 3/2 $g_{0}$ is the dominant one.  Following this hypothesis, the presence of half--integer frequencies would be a clear sign of a period-doubling bifurcation in gravity mode in HH\,Puppis. Such phenomenon had been already reported by \cite{Buchler92} in fundamental radial pressure mode in Pop II Cepheids. 

\subsection{High-order overtones pressure modes}\label{}
A detailed analysis reveals that the power spectrum has a complex structure in the frequency domain around $f_{0}$ and $g_{0}$. After prewhitening of the dominant radial frequency$f_{0}$ and the  gravity frequencies   and their harmonics, we see many additional peaks in the Fourier spectrum (fig.\,\ref{frequency}). We clearly detect three independent frequencies, (1) $f_{2}$ = 4.211334\,$d^{-1}$$\pm$0.00002 with a period ratio $f_{0}$/$f_{2}$ to the main pulsation period $P_{0}$   of 0.61 and an amplitude ration $A_{0}$/$A_{2}$ of 81.46 (2) $f_{1}$=4.604627\,$d^{-1}$$\pm$0.00002, with a period ratio $f_{0}$/$f_{1}$ to the main pulsation period $P_{0}$ of 0.56 and an amplitude ration $A_{0}$/$A_{1}$ of 92.60 (3) 
$f_{3}$=5.921856\,$d^{-1}$$\pm$0.00002, with  a period ratio $f_{0}$/$f_{3}$ to the main pulsation period $P_{0}$ of 0.43 and an amplitude ration $A_{0}$/$A_{3}$ of 101.26.  Table\,\ref{tabratio} gives the relevant values of period, frequency  and amplitude ratios. The frequencies $f_{2}$, $f_{3}$, and $f_{1}$ consist of  linear combinations of the fundamental radial mode,  gravity modes and their harmonics. The origin of the frequencies $f_{2}$, $f_{3}$ and $f_{1}$ is a
great challenge. Judging from the period ratio alone, the period ratios  $f_{0}$/$f_{2}$ and $f_{0}$/$f_{3}$ of 0.61 and 0.43 respectively  are very close to those of the second  and third overtones to fundamental radial mode. Thus the frequency $f_{2}$ and $f_{3}$ are  likely related to the second and third radial overtone pressure mode respectively. In fact, the 0.61 period ratio match the second overtone period ratio perfectly in Fig.\,A2 in \cite{Nemec11}. Furthermore, such period ratio has been observed  in multiple RR\,ab stars, giving more credence to the notion that the second overtone may become excited with low amplitude in these stars (\cite{Lazlo17}). 
The 0.43 ratio is borderline match the calculated OGLE  third overtone RR\,Lyrae  0.45--0.48 range in Fig.\,7 in \cite{Smolec16}. On the other hand, the period ratio of $f_{1}$ to the fundamental radial mode is 0.56. Such an unexpected period ratio of 0.56 might  be interpreted as a nonradial mode.  The additional frequencies 3/2 $f_{3}$ = 8.88293\,$d^{-1}$ might be the third half--integer frequency of the third radial overtone pressure mode. A new group of double periodic stars, in the OGLE Galactic buldge photometry, was detected with period ratios 0.68--0.72. The nature of such additional periodicity still remain unknown (\cite{Prudil17} and \cite{Smolec16}). On the other hand, \cite{Net19} and \cite{Netzel15} detected an additional period, longer than the expected period of the radial first--overtone  mode in OGLE RR\,c stars and concluded that it is not obvious connecting such addition periodicity  to a gravity or  mixed mode.

\subsection{Blazhko modulations and nonradial pressure modes}\label{}
\begin{figure*}[ht!]
\epsscale{0.7}
\plotone{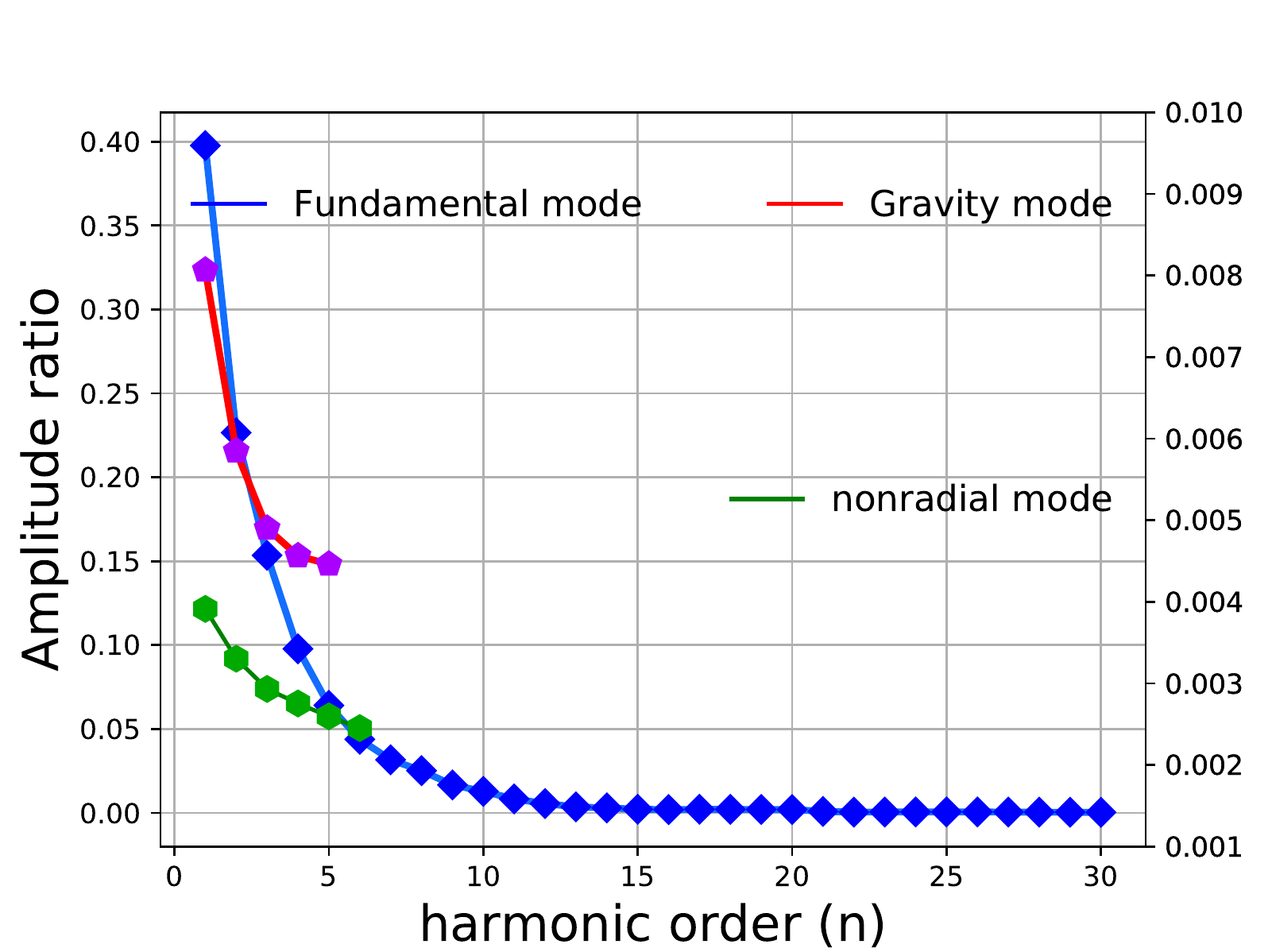}
\caption{ Amplitude ratios of the harmonic components of the fundamental radial pulsation, the gravity mode and the nonradial pressure pulsation}
\label{order1}
\end{figure*}

Besides the nonradial pressure mode  $f_{1}$ = 4.604627\,$d^{-1}$, we detect the independent frequency  $h_{0}$ = 3.043605\,$d^{-1}\pm$0.00002. This frequency appears at  a period  ratio $f_{0}$/$h_{0}$ to the main pulsation period $P_{0}$ of 0.84  with harmonics up to the 6th order (fig.\,\ref{frequency}). Figure\,\ref{order1} demonstrates that the $h_{0}$ harmonic frequency amplitude ratio follow a low quasi--exponential decrease. The frequency $h_{0}$ consists of linear combinations of the fundamental, multi-overtones and gravity modes and their harmonics (Table\,\ref{frequency}). We interpret $h_{0}$ as a nonradial pressure mode in HH\,Puppis. \\ 

\begin{figure}
\gridline{\fig{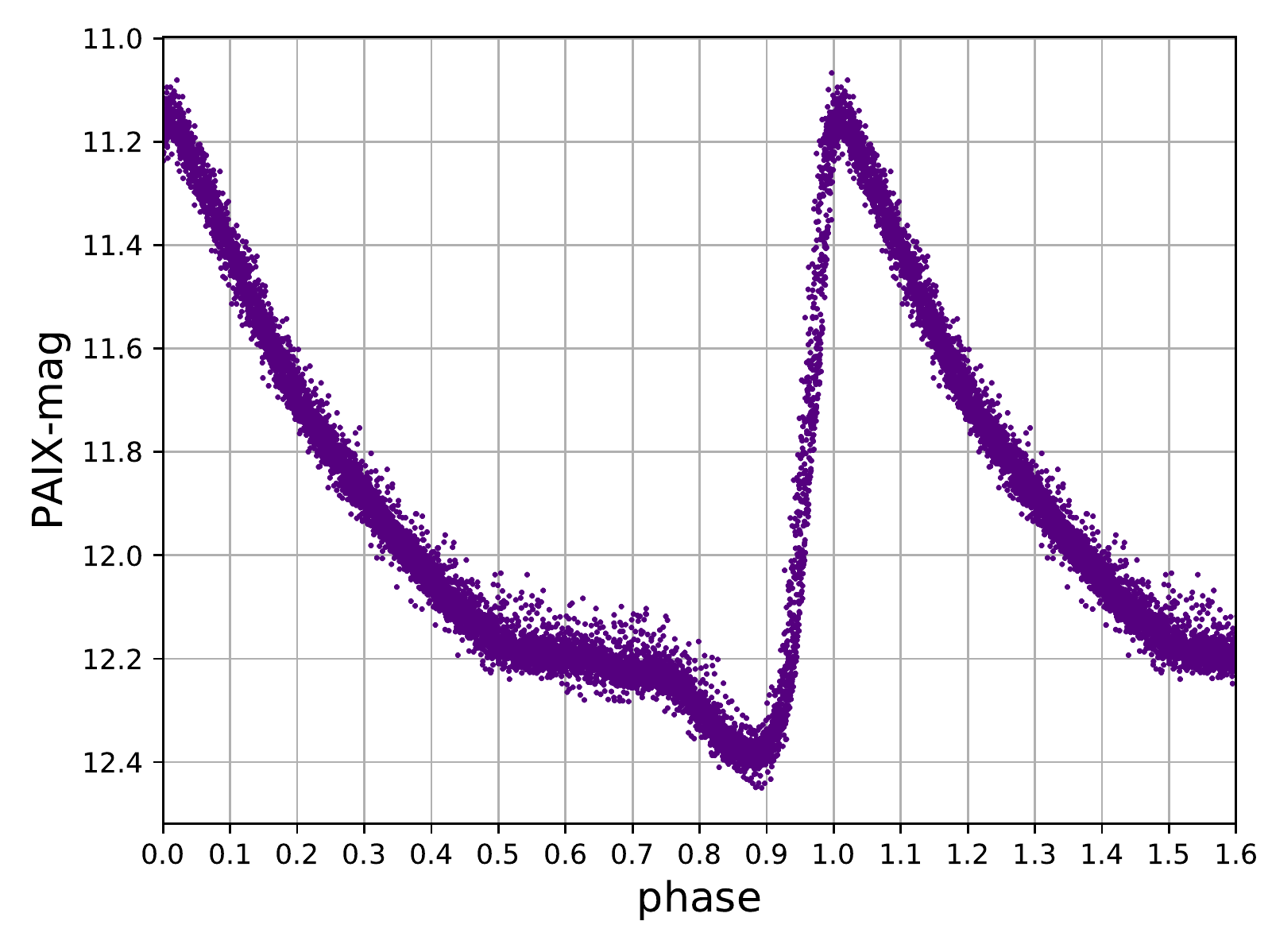}{0.5\textwidth}{(a)}}
\gridline{\fig{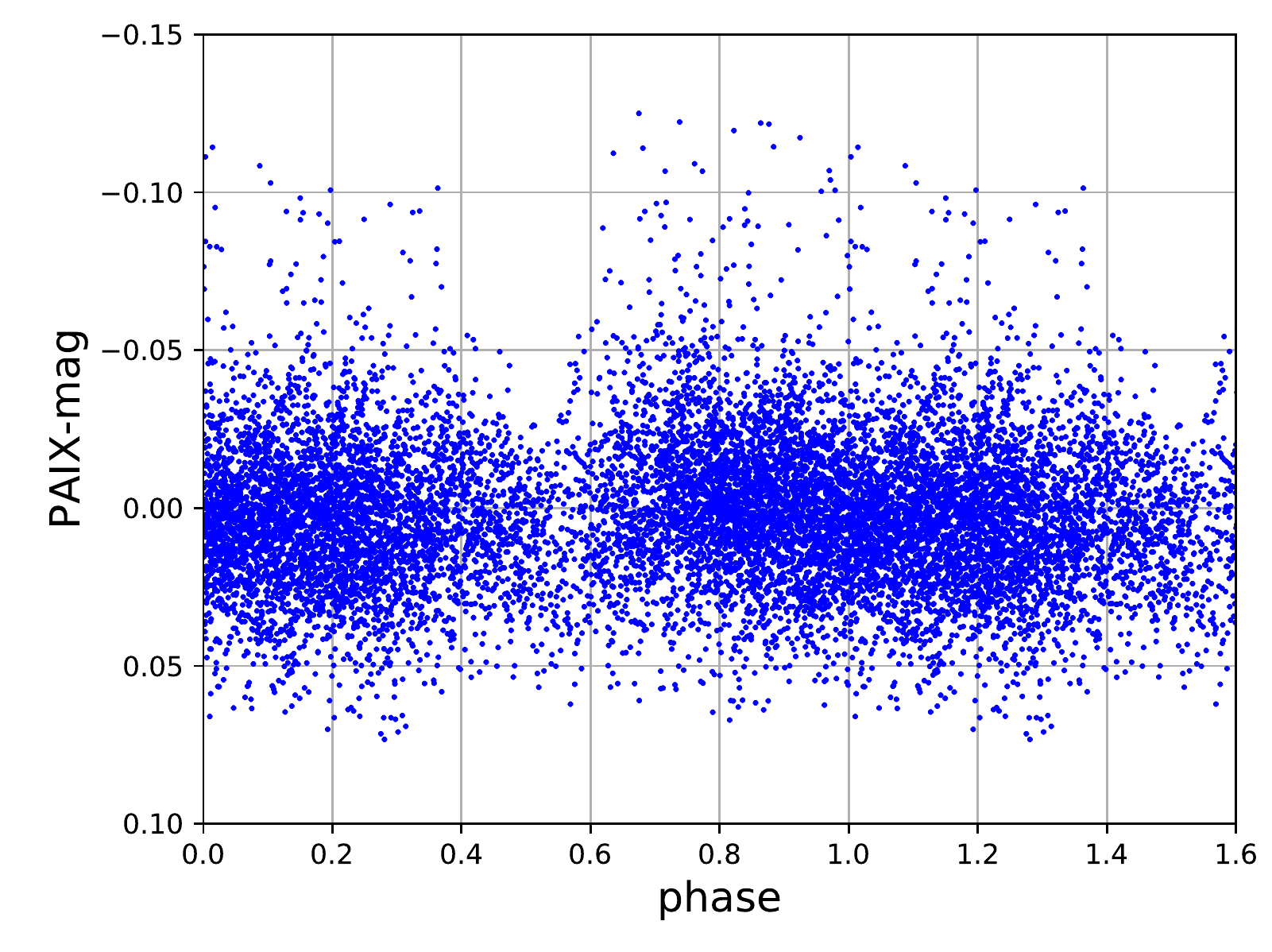}{0.5\textwidth}{(b)}}
\gridline{\fig{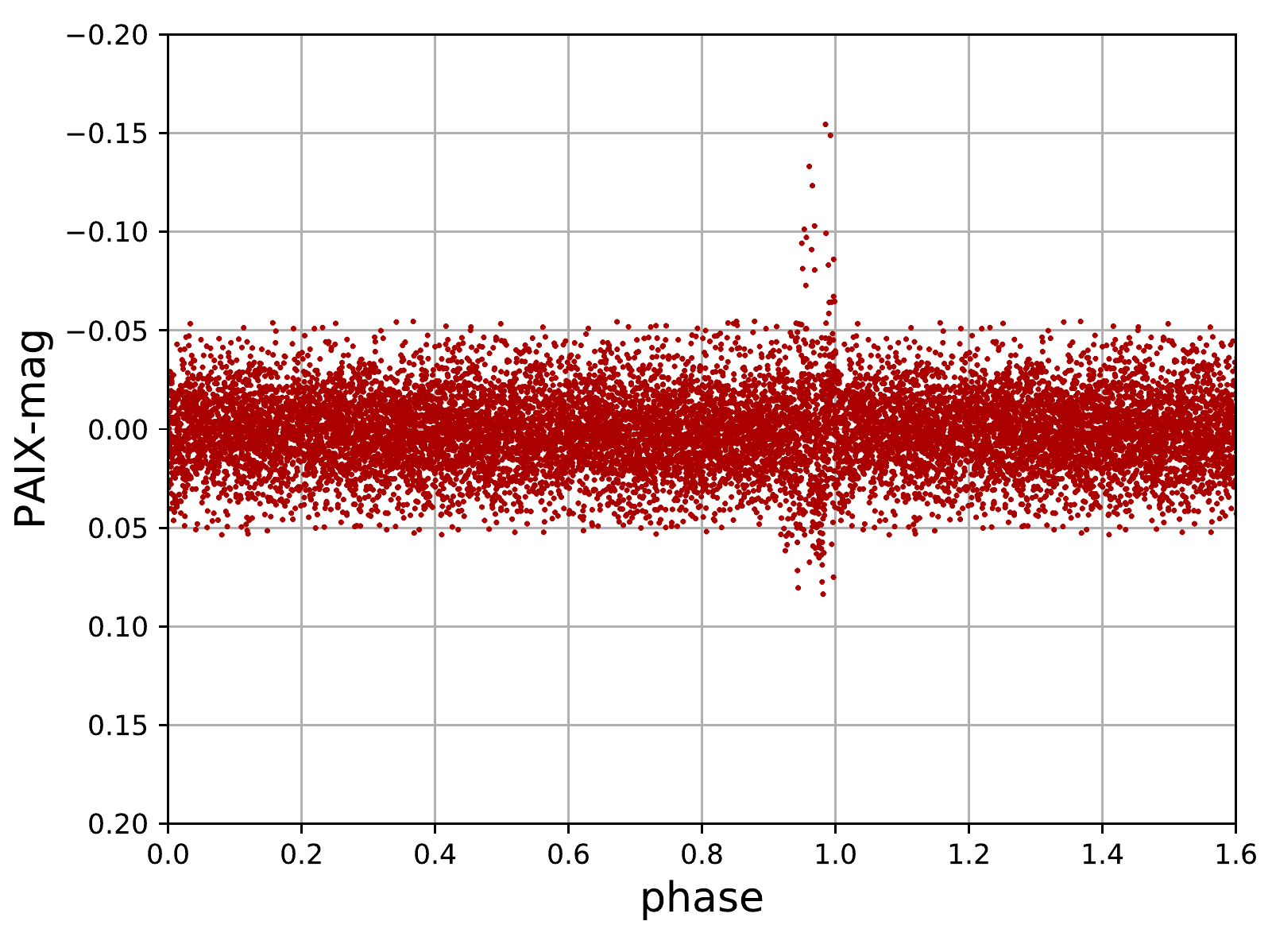}{0.5\textwidth}{(c)}}
\caption{ Folded PAIX light curve of RR Lyrae star HH Puppis. From the upper to lower panel : 
(a) folded with the fundamental period $P_0$; (b) folded with the fundamental period $P_0$ after removal of $P_0$ and harmonics, its multi--overtones pressure modes,  their combination frequencies, Blazhko modulations, and the additional nonradial pressure mode frequency  and harmonics; (c) folded with $P_0$ after removal of all  detected frequencies shown in Table\,\ref{frequency}.\label{LC}}
\end{figure}

\begin{figure*}[ht!]
\plotone{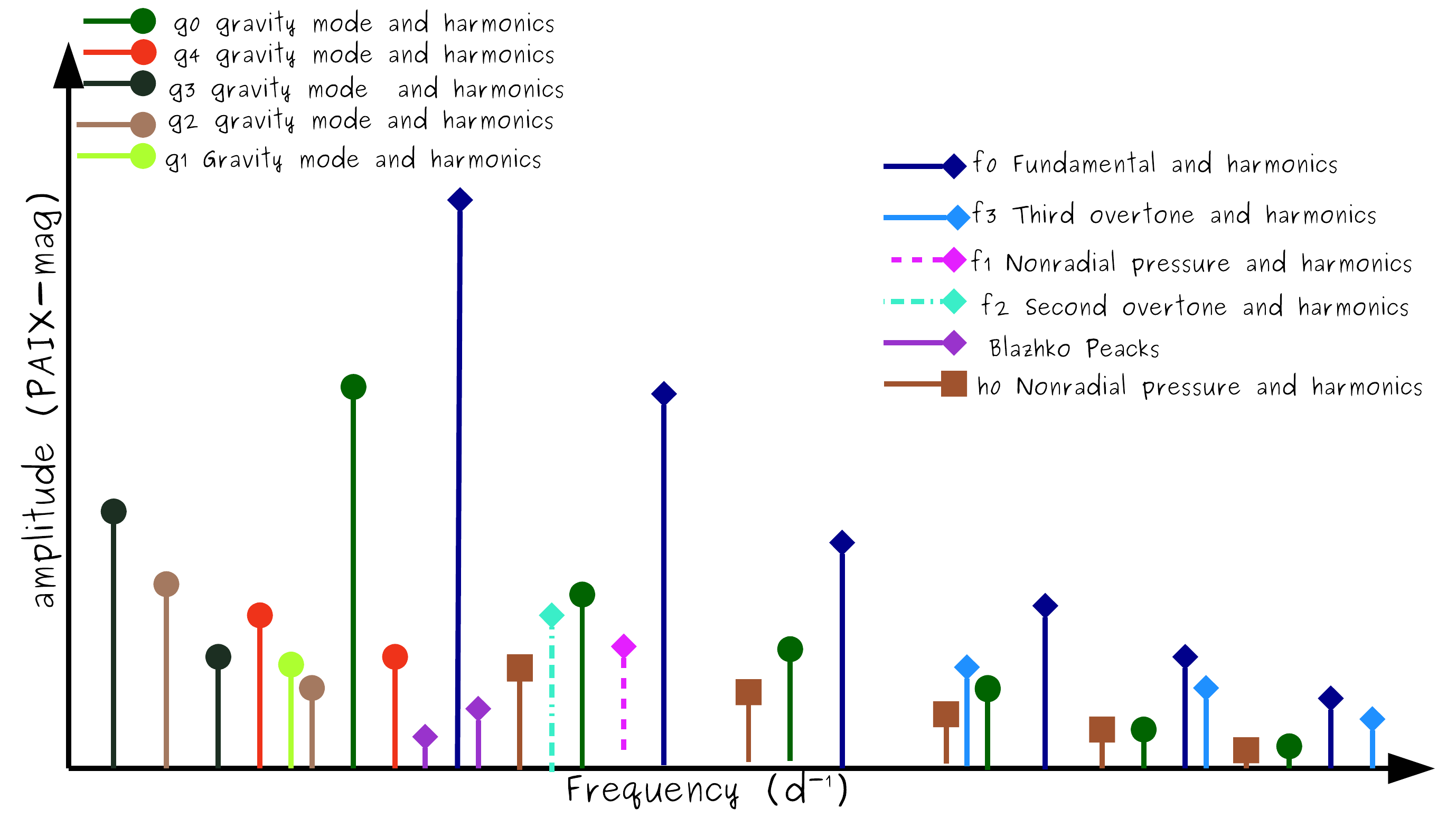}
\caption{Illustration of a complex frequency structure of multiple radial \& nonradial pressure  and gravity  modes, their harmonics \& sub--harmonics, and modulation peaks in HH\,Puppis. }
\label{order}
\end{figure*}

From a purely mathematical point of view, when a  periodic signal is modulated both in the amplitude and phase, the Fourier spectrum shows an infinite series of side peak structures with asymmetric amplitudes between the higher and lower frequency sides. The frequency analysis of HH\,Puppis in this study shows a modulation multiple structure $mf_{0}\,\pm\,kf_{B}$. As shown in Fig.\,\ref{frequency}  a triplet structure $f_{0}$$\pm$$f_{B}$ is clearly detected, showing a Blazhko frequency $f_{B}$ = 0.031039\,$d^{-1}$ (Blazhko period of 32.22\,d). We detect also a quintuplet structure $f_{0}$$\pm$$2f_{B}$ and septuplet structure  $f_{0}$$\pm$$3f_{B}$ and their harmonics as well (Table\ref{frequency}). The  multiplet side peaks (left and right) show an asymmetry. Such asymmetry is caused by the amplitude difference between right and left side peaks. The right--side peaks have a higher amplitude than the corresponding left--side ones. We do not directly detect $f_{B}$ as an independant frequency in the frequency spectrum. This explains the weak behaviour of the Blazhko effect in HH\,Puppis shown in Fig\,\ref{LC}. Finally, in defiance of an extensive literature claiming that HH\,Puppis is a non Blazhko star, we put into evidence that HH\,puppis is a bona-fide Blazhko star. Figure\,\ref{order} gives an illustration of a complex frequency structure of different radial and nonradial  pressure and gravity oscillating modes in HH\,Puppis.\\

 Finally, Figure\,\ref{frequency}\,c shows the residual spectrum prewhitening of all
the frequencies that we identified. To check our frequency solution we ran the program package SigSpec that calculates the spectral significance defined by \cite{Reegen07} for each frequency of the spectrum. We use the sig = 5 value as an over--all criterium. The spectral significance value is given in 
Table\,\ref{tabratio}.

\begin{table}[ht]
 \caption{Summary of the relevant period ratios to main pulsation period and amplitude ratios of HH\,Pup}
 \begin{center}
 \begin{tabular}{||ll|l|c|c|c|c|c|c|c|c|c|}
 \hline
 \hline
 \hline
\bf frequency & & $g_{4}$ &  $g_{3}$ &  $g_{2}$ & $g_{1}$ & $g_{0}$ & $f_{0}$ & $f_{1}$ & $f_{3}$ & $f_{2}$ &$h_{0}$\\
 \hline
 \hline
 
 \bf spectral significance & &  210.90 &  270.45 & 215.78 &110.92& 350.12 & 2910.88&180.97&140.63&198.50 & 135.37\\
 \hline
 \hline

\bf period ratio & &  2.08 &  3.03 & 2.22 &1.72& 1.28 & 1&0.56&0.43&0.61 & 0.84\\
 \hline
 \hline
\bf amplitude ratio & &  72.72 & 58.31 & 67.96 &125.41 & 49.32&1& 92.60&101.26&81.46& 101,66\\
   \hline
   \hline
    \end{tabular}
 \label{tabratio}
 \end{center}
 \end{table}

\section{Fundamental parameters}\label{parameters}

\begin{figure*}
\gridline{\fig{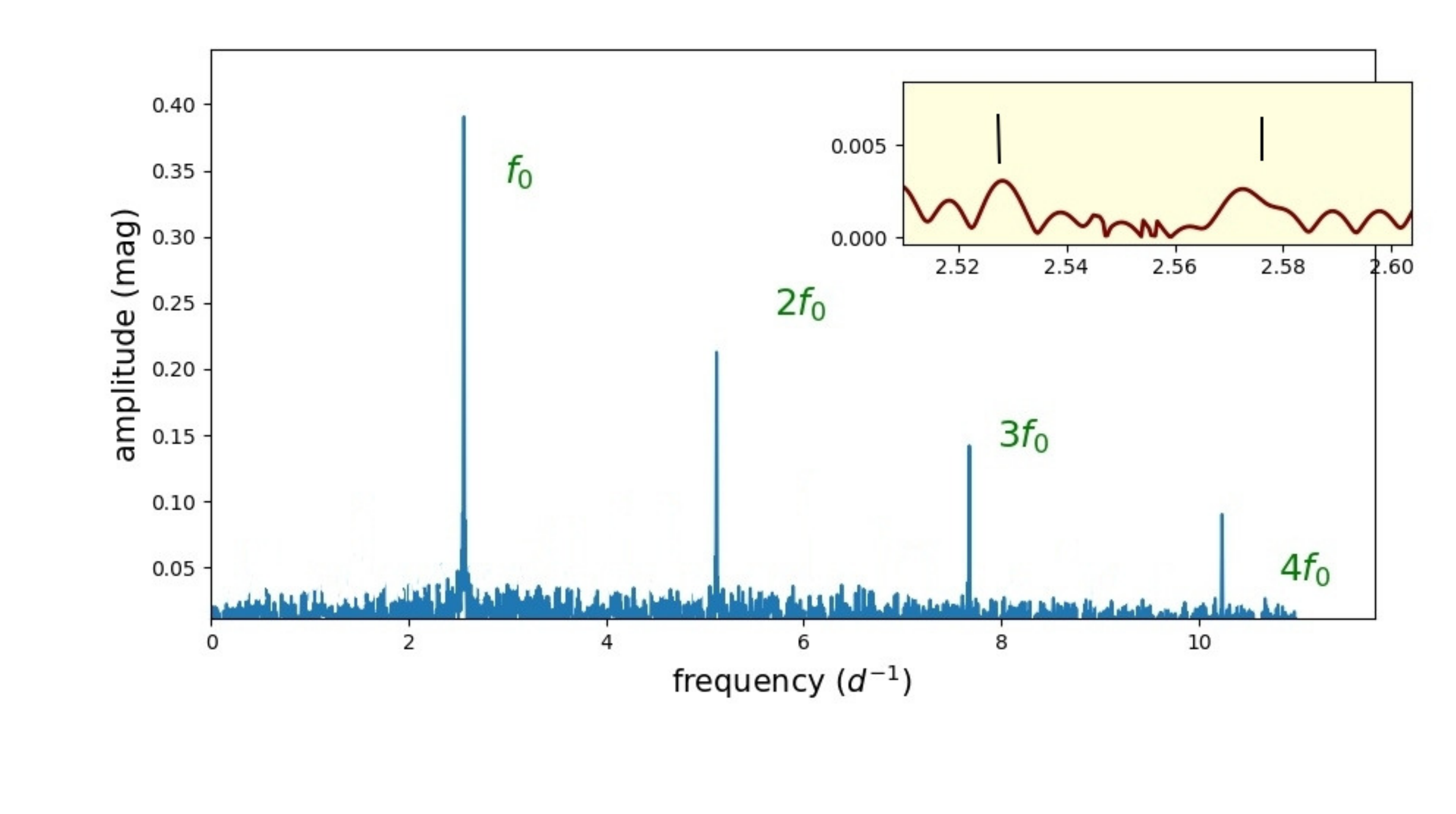}{0.8\textwidth}{(a)}}
\gridline{\fig{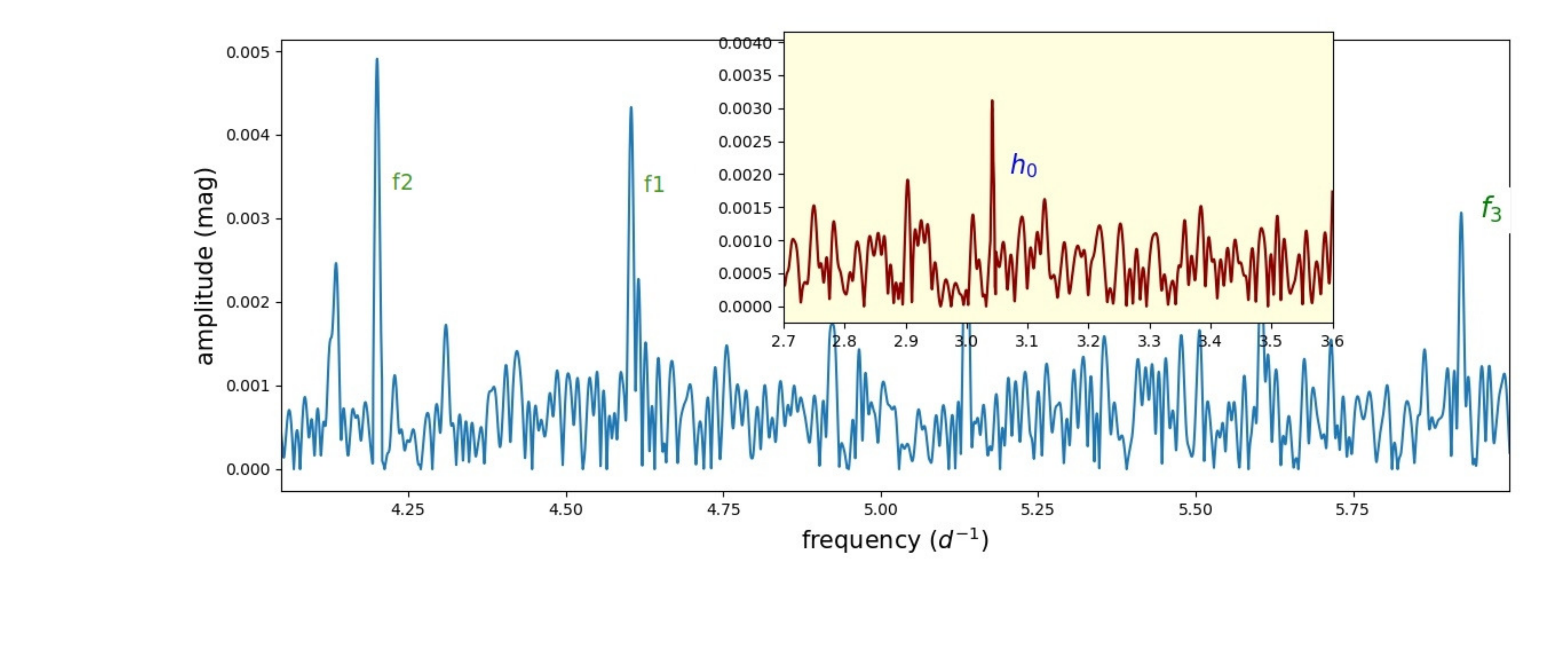}{0.8\textwidth}{(b)}}
\gridline{\fig{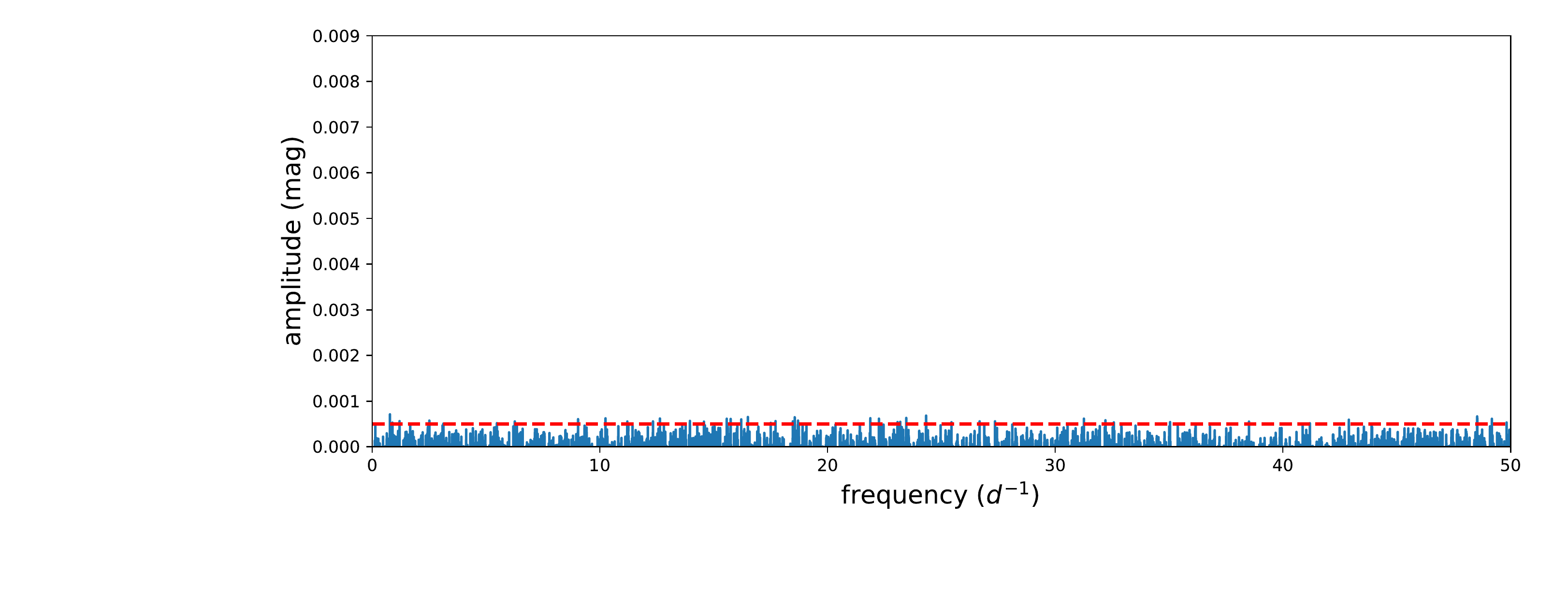}{0.8\textwidth}{(c)}}
\caption{From the upper to lower panels: (a) the amplitude spectrum of the PAIX light curve of RR Lyrae star HH Puppis;  The insert panel shows the modulated peaks $f_{0}\,\pm\,f_{B}$. (b) after prewhitening the fundamental radial frequency, the nonradial gravity modes, their harmonics and their linear combination terms; The insert panel shows the nonradial frequency $h_{0}$ and (c) 
the residuals after prewhitening with all frequencies in Table\,\ref{frequency}. The dashed curve shows the noise level that corresponds to the level of the weakest amplitude in the data. \label{frequency}} 
\end{figure*}

A good synthesis of the available formulae to compute physical characteristics of RR\,Lyrae stars can be found in \cite{Nemec11}. We use here Sandage methods \cite{Sandage04} to estimate metal abundances:
\begin{itemize}
 \item using the period--amplitude relation
 \begin{equation}
  [Fe/H] = -1.453 A_{0} - 7.990 logP - 2.145
 \end{equation} 
We obtain a [Fe/H] value of $-0.53$$\pm$0.043\,dex. The coefficients have respectively uncertainties of $\pm$0.027 and $\pm$0.091 respectively (from \cite{Sandage04}).
\end{itemize}
\begin{itemize}
 \item the period--rise--time relation
 \begin{equation}
 [Fe/H] = 6.33 {RT} - 9.11 logP - 4.60
 \end{equation}
 Where RT is the rise--time, giving a [Fe/H] value of $-0.25$\,dex. 
 \end{itemize}
  from the period--amplitude relation and the period--rise--time relation respectively.
 \begin{itemize}
 \item the period-phase relation
 \begin{equation}
 [Fe/H] = 1.411 {\Phi}_{31} - 7.012 logP - 6.025
 \end{equation}
where ${\Phi}_{31}$ is the Fourier phase parameters for a cosine
version which differ by $\pi$ radians with the one derived from a
sine version : $\phi^{c}_{31}$ = $\phi^{s}_{31}$ - 3.14159 (using ’s’ and ’c’ superscripts
for phase parameter computed with, respectively, sine and cosine series). The ${\Phi}_{31}$ and the $logP$ coefficients have respectively uncertainties of $\pm$0.014 and $\pm$0.071 (from \cite{Sandage04}). We derive a [Fe/H] value for a cosine version of 4.51 $\pm$0.078\,dex. 
 \end{itemize}
As discussed by \cite{Nemec11}, systematic discrepancies are seen between each of these 3 methods. We adopt the results from the period--amplitude method with a [Fe/H] value of $-0.53$$\pm$0.043\,dex. The [Fe/H] value derived from the period--phase relation is very high. Such issue is due to the very high number of frequency components that induces an overfitting and then altering the phase values. Moreover, we state that our fundamental parameters study is generated, as described in Section\,\ref{data}, by use of the V--filter passband ($\lambda_{eff}$\,=\,540\,nm and $\Delta\lambda$\,=\,100\,nm)  that is very narrower than the  Kepler spectral passband (400\,nm -- 850\,nm) which Nemec's results have been inferred. High-resolution echelle spectra from our previous HH\,Puppis studies give a [Fe/H] spectroscopic value of -0.95\,dex (\cite{Preston19}) and  -0.69\,dex (\cite{Chadid17}). A recent spectrocopic estimation of the HH\,puppis [Fe/H] shows a value of -0.73\,dex (\cite{Crestani21}. Such high--spectroscopic studies compare very favourably with the estimated $-0.53$\,dex period--amplitude method with the other methods. \\

We calculate the dereddened mean B-V colour from \cite{Walker01}
\begin{equation}
(B - V)  = 0.189 logP - 0.313 A_{1} + 0.293 A_{3} + 0.460
\end{equation}
\begin{equation}
logg = 2.473 - 1.226 logP
\end{equation}
\begin{equation}
log T_{eff} = 3.8840 - 0.3219(B - V ) + 0.0167 logg + 0.0070 [Fe/H]
\end{equation}

We infer a $B-V$ value of $0.303\,mag$, a mean surface gravity of $2.971$ wich is higher than the high  spectroscopic values listed in \cite{Chadid17}, and an effective temperature value of 6790$^{\circ}$K. \\
Following \cite{Caputo00} stellar evolution models, we calculated the 
absolute magnitude, Pulsation luminosity and mass.
\begin{equation}
M_{V} = 0.18 [Fe/H] + 1.05
\end{equation}
giving a $M_{V}$ value of 0.95$\pm$0.01 (the uncertanties are from \cite{Nemec11}.
\begin{equation}
logL = 1.538 - 0.110 [Fe/H]
\end{equation}
showing a $L$ value of 39.5
\begin{equation}
logM = -0.283 - 0.066 [Fe/H]
\end{equation}
inferring a mass value of 0.57\,$M_{\odot}$.\\
Table\,\ref{tabfun} summarizes the main results of HH\,puppis fundamental parameters.

\begin{table}[ht]
 \caption{Observational characteristics of HH\,Pup}
 \begin{center}
 \begin{tabular}{|l|c|c|c|c|c|c|c}
 \hline
 \hline
 $\bf [Fe/H]$ &  {$\bf (B-V)$} &  {$\bf logg$} &   $\bf T_{eff}$ & $\bf M_{V}$ &  {$\bf L$} & {$\bf M$}  \\ 
 {dex} & {mag} &  &  $^{\circ}$ K &  &  & $M_{\odot}$ \\ 
 \hline
 \hline
{\bf -0.53$\pm$0.043} & {\bf 0.303} & {\bf 2.971} & {\bf 6790} & {\bf 0.95$\pm$0.01} & {\bf 39.5} & {\bf 0.57}\\
   \hline
   \hline
    \end{tabular}
 \label{tabfun}
 \end{center}
 \end{table}
 
\section{Shock waves and dynamics}\label{shock}
As described by \cite{Chadid14}, shock waves induce a high distortion in the RR\,Lyrae Light Curve (fig.\ref{fig}). Multiple bumps were observed by \cite{Chadid14} and interpreted as a multi-shock structure in the RR\,Lyrae atmosphere. More  recently, \cite{Prudil17} have determined the centers and strengths of main and early shock features in the phased fundamental--mode RR Lyrae light curves.
Figure\,\ref{LC} shows the residual light curve after prewhitening with the whole detected frequencies in Table\,\ref{frequency}, in an attempt to remove all excited oscillating modes from the data set. The most striking
feature of the residual light curve is the intense variation
that appears around phase 0.95 in a narrow 8\% phase interval of the pulsation cycle. Such variation induced by a strong shock weave, named $Sh_{H+He}$ by \cite{Chadid14}, created by the $\kappa$ and $\gamma$ mechanisms, traversing the photosphere of HH\,Pup during the rise time of the light curve. Such strong shock wave is at the origin of hump observed just before the light curve maximum
during successive pulsation cycle (\cite{Chadid06}) and (\cite{Chadid13}). A similar residual scatter had been already observed at the light curve maxima of the Blazhko stars S\,Arae occurring in a phase interval of 10\% of the pulsation period.  DR\,Andromeda, RR\,Lyr  and  RR\,Gem  show a similar residual scatter as well, occurring in a phase interval of 20\% of the pulsation cycle. This means that the variation in the main shock strength is higher in HH\,puppis than in those stars.

\section{Excitation mechanism of gravity waves in RR\,Lyrae stars}\label{gmodes} 
Our results show that  HH\,Puppis is a metal--rich Blazhko RR\,ab star. According to Chadid's classification (\cite{Chadid17}) HH\,Puppis is a hypersonic--regime RR\,ab star showing a largest gravity acceleration\footnote{The term dynamical acceleration is associated to the primary acceleration induced by the  $Sh_{H+He}$ primary shock. The term gravity acceleration  is associated to the secondary acceleration and the term gravitational acceleration is log\,g.}, its dynamical atmosphere shows the highest dynamical acceleration and  the greatest $Sh_{H+He}$ main shock Mach number. The most striking finding in this study is that HH\,Puppis has an even richer Fourier spectrum,  and undergoes at the same time  $p$ modes and $g$ modes. Such simultaneity of gravity and pressure  modes   results from the coupling of pressure waves that probe the atmosphere and gravity waves that probe the radiative core giving a direct access to the core of HH\,Puppis. \\ 

The detection of any $g$ mode signals in the Sun, let alone in solar--like stars,  is still a highly contentious issue today. The major argument consisted in the difficulty in proving a possible mechanism for $g$ modes excitation. The pressure modes propagate at high frequencies through the convective zone up to the surface, while gravity modes are trapped at low frequencies in the radiative interior. In the convection zone, hence outside the radiative region, the gravity waves are evanescent. The $g$ modes has been predected and detected in Solar--type stars. They were a treasure trove for red giant asteroseismology as described by \cite{bed11} and \cite{bec11}. Both studies pointed out that the $g$ modes are nonradial modes.\\
In RR\,Lyrae stars, \cite{Van98} and \cite{Jumbo16} theoretically predicted that there is no gravity wave cavity for radial modes and that the evanescent between the $g$ and $p$ mode cavities is narrower for $l$ = 2, 3 and 4 where strong mixing can be expected. Moreover, the driving rates drop dramatically for $l$=1 modes below the fundamental mode frequency (\cite{Jumbo16}). On the other hand, the internal gravity waves are an efficient transport mechanism in stellar radiative zone (\cite{Schatzman93}). Strong downward plumes induce a substantial distance extension into the adjacent stable zones then the internal gravity waves are randomly generated  (\cite{Hurlburt86}). The downward plumes are born in  the upper envelope convective layers, are accelerated by the buoyant force and finally end their life in the overshoot region after transferring a large amount of their stored kinetic energy to the stably stratified medium resulting in an internal gravity wave field. This convective penetration  strongly depends on the value of the local P\'eclet number (\cite{Zahn91} and \cite{Dintrans05}). The depth of the penetration in the stably stratified zone is a crucial parameter in the excitation mechanism. Stronger the P\'eclet number, thinner the  penetration and lower gravity waves excitation. When the P\'eclet number is larger the plume is rapidly stopped by the buoyancy breaking. In the opposite case, the P\'eclet number is smaller implies that the buoyancy breaking rapidly disappears, leading to a large penetration and higher  excitation of the internal gravity waves.  We assume that the convective envelope in RR\,Lyrae stars is larger enough to create a small local P\'eclet number which induces an adequate penetration power that is sufficiently able to excite the very small $g$ mode amplitudes,  as detected in this study. Accordingly, we may hypothesize that the excitation of internal gravity waves in RR\,Lyrae stars is possible by the penetration of convective plumes into the adjacent stably stratified radiative zone. It is worth highlighting that the treatment of convective contributions on stellar stability in RR\,Lyrae stars is a serious problem and has been debated for decades. As recently suggested  by \cite{Ste13}, the convective envelopes of RR\,Lyrae stars are deeper. \cite{Huet10} found that shocks encountering the density inhomogeneities characteristic of convective envelopes  generate turbulent velocities that dissipate shocks. Thus, he  attributed such shock energy dissipation to  larger convective envelopes in RR\,Lyrae stars. More recently, \cite{Chadid17} detected that the photospheric radii and the radius variations of RR\,Lyrae stars are larger and are a direct consequence of their greater thicknesses of the  compression zones. In that way,  they concluded that RR\,Lyrae stars are larger convective envelopes. However, the understanding of process associated with convection in RR\,Lyrae stars makes our hypothesis on the  mechanism of gravity waves excitation  a challenging endeavor. Conclusively, the RR\,Lyrae pulsation is excited by several distinct mechanisms, the $\kappa$--$\gamma$  mechanisms and the radiation--modulated excitation mechanism that induce  pressure waves, and the penetrative convection mechanism  that induces  gravity waves. We hypothesize that the  key process in the Blazhko mechanism is the gravity waves that play a trigger role. The pressure waves, induced by the $\kappa$--$\gamma$ mechanisms, are perturbed by the gravity waves, induced by the convection mechanism, leading to cyclic modulations,  the so--called Blazhko effect. Additional theoretical efforts will be addressed in our theoretical subsequent paper focused on a numerical modeling towards a better understanding of these issues.

\section{Summary}\label{con}
We investigate unprecedented time--series ground--based observations, by use of  PAIX in the highest plateau of Antarctica over 1 polar night, and an extensive frequency analysis leading  us to some important conclusions. The first and by far the most important one is the detection of the gravity modes in RR Lyrae stars. The power spectrum of HH\,Puppis shows a complex frequency structure, in particular dominant peaks occur at low frequencies showing nonradial gravity modes. These lower frequencies and harmonics linearly interact with the dominant fundamental radial pressure mode and its second and third overtone pressure modes. Half--integer frequencies of $g$ modes are detected, likewise side peak structures  demonstrating that HH\,puppis is a bona--fide Blazhko star. \\

We interpret the excitation mechanism of internal gravity waves in RR\,Lyrae stars by the penetrative convection. We demonstrate that RR\,Lyrae stars undergo several distinct driving mechanisms, (1) the pressure oscillating excitation mechanisms  $\kappa$ and $\gamma$ mechanisms occur in hydrogen and helium innization zones, inducing the outward radiative shock  $Sh_{H+He}$, (2) the pressure oscillating excitation mechanism, radiation--modulated excitation mechanism, occurs in a zone of the radiation flux gradient which is the bottom and the top of the convective zone, inducing the outward radiative shock $Sh_{RME}$, and (3) the gravity oscillating excitation mechanism, convection mechanism  occurs just below the photosphere. \\

Finally, we hypothesize that the Blazhko effect is a nonlinear gravitohydroynamical interplay between the gravitation waves, that play the trigger role, intimately connected to the convective zone right bellow the photosphere, and the pressure waves provided by variations of the opacity  in the $H$ and $He$ ionization zones, $\kappa$ and $\gamma$ driving mechanisms.

\section{Acknowledgments}
 
The PAIX project has been  supported by the United States Air Force Research Laboratory through the European Office of
Aerospace Research and Development -- US Air Forces F61775-02-C002  --\\
I am grateful  to Jean Vernin and George Jumper from Hanscom Air Force Base,  for their friendly and continuous multilevel support.  I acknowledge all  the explorers and the scientific researchers who have contributed to make my polar expeditions and Antarctica projects  less bumpy rides  under  extreme conditions of Antarctica.

\bibliography{HHPup}{}
\bibliographystyle{aasjournal}

\clearpage
\begin{longtable}{cccc}
\hline 
\\
 {\bf Frequency} & {\bf Amplitude} & {\bf Phase} & {\bf ID} \\
{$\mathrm{[d^{-1}]}$} & {[mag]} & {[cycles]} &  \\ 
\\
\hline 
\\
\multicolumn{4}{l}{{\bf A- Main fundamental pressure mode frequency and harmonics}} \\
\\
2.559196&	 0.397679 &	 0.501680 &    $f_{0}$ \\
5.118375&	 0.226581 &	 0.162268 &    $2f_{0}$ \\
7.677543&	 0.153467 &	 0.870234  &    $3f_{0}$\\
10.236743&	 0.097789 &	 0.728544 &    $4f_{0}$\\
12.795857&	 0.064069 &	 0.215091 &    $5f_{0}$\\
15.355083&	 0.043846 &	 0.007674&    $6f_{0}$ \\
17.914130&	 0.031658 &	 0.866588 &    $7f_{0}$\\
20.473334&	 0.025181 &	 0.917582 &    $8f_{0}$\\
23.032741&	 0.016695 &	 0.235621 &    $9f_{0}$\\
25.592196&	 0.012829 &	 0.884852&    $10f_{0}$ \\
28.151083&	 0.008318 &	 0.248028&    $11f_{0}$ \\
30.710335&	 0.005599 &	 0.555862&    $12f_{0}$ \\
33.268002&	 0.003571 &	 0.469407 &    $13f_{0}$ \\
35.828500&	 0.003000 &	 0.480355 &    $14f_{0}$\\
38.387956&	 0.002193 &	 0.851552 &    $15f_{0}$\\
40.947905&	 0.001961 &	 0.981247 &    $16f_{0}$\\
43.506891&	 0.002156 &	 0.241041 &    $17f_{0}$\\
46.066245&	 0.002121 &	 0.348781 &    $18f_{0}$\\
48.625231&	 0.002034 &	 0.753122&    $19f_{0}$\\
51.181787 &	 0.002055 &	 0.396982 & $20f_{0}$\\
52.995878 &	 0.000902 &	 0.441381 & $21f_{0}$\\
55.994399 &	 0.000510 &	 0.385783 & $22f_{0}$\\
57.995152 &	 0.005274 &	 0.387343 & $23f_{0}$\\
60.995253 &	 0.000573 &	 0.722048 & $24f_{0}$\\
62.997290 &	 0.000596 &	 0.750047 & $25f_{0}$\\
66.004626 &	 0.000603 &	 0.958771 & $26f_{0}$\\
68.996565 &	 0.000486 &	 0.808693 & $27f_{0}$\\
70.994377 &	 0.000435 &	 0.447202 & $28f_{0}$\\
73.988838 &	 0.000372 & 	 0.336705 & $29f_{0}$\\
76.001869 &	 0.000324 &	 0.556226 & $30f_{0}$	\\ 
\\
\multicolumn{4}{l}{{\bf B- Second and Third overtone mode frequencies, harmonics and sub--harmonics}}\\
\\
4.211334	 & 0.004882 &	 0.622764 & $f_{2}$	\\
5.921856	 & 0.003927 & 	 0.200287 & $f_{3}$	\\	
11.840311	&  0.004298 & 	 0.968200 & $2f_{3}$\\
8.88293	 & 0.003954 & 	 0.675503 & $3/2f_{3}$	\\ 
17.766750 	& 0.003612 &	 0.738022 & $3f_{3}$	\\
\\
\multicolumn{4}{l}{{\bf C- Gravity mode frequencies, harmonics and sub--harmonics}} \\
\\
2.001301&	 0.008064 &	 0.937279 &    $g_{0}$\\ %
4.003510&	 0.005846 &	 0.954913 &   $2g_{0}$\\
6.005067&	 0.004900 &	 0.190098 &    $3g_{0}$\\
8.006805&	 0.004562 &	 0.541087&    $4g_{0}$\\
10.005108&	 0.004463&	 0.784127 &    $5g_{0}$\\
1.0006743&	 0.001978 &	 0.124666  &    $1/2g_{0}$\\
3.001893&	 0.007024 &	 0.295939  &    $3/2g_{0}$\\
5.003268&	 0.002896 &	 0.200007  &    $5/2g_{0}$\\
7.004667 &	 0.002469 & 	 0.055471 & $7/2g_{0}$	\\
0.835433&	 0.006820 &	 0.971325 &    $g_{3}$\\ 
1.670510&	 0.003702 &	 0.381325&  $2g_{3}$ \\
1.150662&	 0.005852 &	 0.624346 &    $g_{2}$\\ 
2.301243&	 0.003946 &	 0.041390 &  $2g_{2}$\\
1.233302&	 0.005469 &	 0.562601&   $g_{4}$  \\      
2.466715&	 0.003412 &	 0.222470&  $2g_{4}$\\
1.498787&	 0.003171 &	 0.834126  &   $g_{1}$\\ 
2.997876&	 0.001578 &	 0.018760&  $2g_{1}$\\
\\
\multicolumn{4}{l}{{\bf E-  Blazhko Multiplet frequencies}} \\
\\
2.528157&	 0.003964&	 0.967529 &  $f_{0}-f_{B}$\\
2.590235&	 0.005161&	 0.367529 &  $f_{0}+f_{B}$\\
2.497206&	 0.003196 &	 0.529648 &  $f_{0}-2f_{B}$\\
2.621230&	 0.004440 &	 0.368648 &  $f_{0}+2f_{B}$\\
2.652499&	 0.003167 &	 0.255279&  $f_{0}+3f_{B}$ \\
2.466099&	 0.002887&	 0.535975& $f_{0}-3f_{B}$ \\
\\
\multicolumn{4}{l}{{\bf F-  Nonradial pressure modes and harmonics}} \\
\\
3.043605&	 0.003912 &	 0.125482 & $h_{0}$\\
6.087216&	 0.003302 &	 0.009045 & $2h_{0}$\\
9.130828&	 0.002931 &	 0.179416 & $3h_{0}$\\
12.174814&	 0.002753 &	 0.419028& $4h_{0}$\\
15.219031&	 0.002550&	 0.865765& $5h_{0}$\\
18.261935&	 0.002421 &	 0.578192& $6h_{0}$\\
4.604627	 & 0.004287 & 	 0.905535 & $f_{1}$	\\
\\
\multicolumn{4}{l}{{\bf G-  Linear combinations}} \\
\\
12.238282 &	 0.004053 &	 0.205536 & $4f_{0}+g_{0}$	\\
 12.76466 &	 0.002540 & 	 0.438686 & $5f_{0}-f_{B}$\\
10.267802 &	 0.003196 & 	 0.333569 & $4f_{0}+f_{B}$\\
 5.561218	  & 	 0.004942 &	 0.649476 & $f_{0}+3/2g_{0}$	\\
 28.181595 &	 0.002912 & 	 0.530506 & $11f_{0}+f_{B}$\\
 4.560559	 & 	 0.005007 &	 0.496673 & $f_{0}+g_{0}$	\\
 6.559175	  & 	 0.002883 &	 0.250258 & $f_{0}+2g_{0}$\\
7.1637440	 & 	 0.003445 &	 0.265586 & $f_{1}+f_{0}$\\
4.529445&	 0.002558 &	 0.481783  & $f_{0}-f{B}+g_{0}$\\
1.599963 &	 0.003358  &	 0.978659 & $2g_{4}-g_{3}-f_{B}$\\
7.962294	 &       0.004340 &	 0.982464 & $2g_{4}+f_{0}+3/2g_{0}-2f_{B}$	\\
10.68000 &	 0.004693 &	 0.596918 & $3f_{0}+3/2g_{0}$	\\
3.792523	 &       0.003350 &	 0.877481 & $f_{0}+g_{4}$	\\
0.397909	 &       0.005559 &	 0.459725 & $g_{4}-g_{3}$	\\
6.794516	 &       0.003678 &	 0.066399 & $g_{4}+f_{0}+3/2g_{0}$	\\
0.257877	  &      0.004355 &	 0.927123 & $f_{0}-2g_{2}$ \\
 0.907092	 &       0.004258 &	 0.666979 & $2f_{0}-f_{2}$	\\
8.910838	  &      0.003131 & 	 0.897412 & $3f_{0}+g_{4}$\\
0.514026	 & 	 0.003989 &	 0.402010 & $f_{1}-2f_{0}$\\
2.836750	  &	 0.002325   &     0.456325 & $g_{0}+g_{3}$	\\
 1.652510	 & 	 0.003702 &	 0.381325 & $f_{2}-f_{0}$\\
9.723002	  & 	 0.002660 &	 0.639489 & $f_{1}+2f_{0}$\\
4.057497&	 0.002214 &	 0.602944 & $f_{0}+g_{1}$\\
1.325947&	 0.003478 &	 0.235078 &    $f_{0}-g{4}$\\
2.014383&	 0.004472 &	 0.964364 &    $f_{1}-f_{0}-f_{B}$\\
3.072640&	 0.002314 &	 0.546870 &  $3f_{0}-f_{1}$\\
11.039989&	 0.003336 &	 0.921603 & $2f_{0}+f_{3}$	\\
14.402945&	 0.003289 &	 0.751847 & $2f_{3}+f_{0}$\\
9.222667	 &       0.003901 &	 0.017743 & $2f_{3}-f_{0}-2f{B}$	\\
 5.045954	  & 	 0.002640 &	 0.943928 & $f_{2}+g_{3}$\\
 6.212879&	 	 0.004764 &	 0.663111 & $f_{2}+g_{0}$\\
 2.209971&	 	 0.004708 &	 0.442860 & $f_{2}-g_{0}$ \\
 3.920347&	 	 0.004708 &	 0.719098 & $f_{3}-g_{0}$ \\
 8.213945&	 	 0.002876 &	 0.711630 & $f_{2}+2g_{0}$\\
 0.208515&	 	 0.008043 &	 0.226481 & $f_{2}-2g_{0}$ \\
 4.180270&	  0.003220 &	 0.196335 & $f_{2}-f_{B}$\\
 5.890860 &	  0.002701 & 	 0.929047 & $f_{3}-f_{B}$	\\
21.218494&	 0.006013&	 0.515658& $5f_{0}+2f_{2}$\\
3.105684&	 0.002758 &	 0.955653& $h_{0}+2f_{B}$ \\
4.276853 &	 0.002923 &	 0.465204 &  $h_{0}+g_{4}$	\\
7.254879&	 0.003575 &	 0.078563&  $h_{0}+f_{2}$ \\
1.167822&	 0.005257&	 0.918952 &  $f_{2}-h_{0}$ \\
7.648301&	 0.003989&	 0.866471&  $h_{0}+f_{1}$\\
1.755697	 &       0.004325 &	 0.650188 & $f_{3}-3f_{0}$	\\
11.562553 &	 0.003138 &	 0.264277 & $5f_{0}-g_{4}$	\\
7.584791	 &       0.003024 &	 0.112463 & $2f_{0}+2g_{4}$	\\
16.588501 &	 0.002601 &	 0.618216 & $6f_{0}+g_{4}$	\\
10.133190  & 	 0.002539 &	 0.215918 & $f_{2}+f_{3}$\\
4.285721  & 	 0.002827 &	 0.887962 & $4f_{0}-f_{3}$\\
3.977686	 &  	 0.002974 &	 0.889681 & $3f_{0}-3g_{4}$\\
14.031970 &	 0.002273 &	 0.884726 & $5f_{0}+g_{4}$\\
\hline
\hline 
\\
\caption{\label{tbl-1}Relevant  Frequencies and their Fourier amplitudes, phases, and their identification. (This Table can be found in its entirety in the electronic version of this journal)}
\label{frequency}
\end{longtable}

\end{document}